\documentclass[twocolumn,prb]{revtex4}
\usepackage{graphicx}
\begin{document}
\title{Transport of Entanglement Through a Heisenberg-XY Spin Chain}
\author{ V. Subrahmanyam$^{(1)}$ and Arul Lakshminarayan$^{(2)}$ }
\address{$^{(1)}$ Department of Physics, Indian Institute of Technology, Kanpur, 208016, India.
\\$^{(2)}$ Department of Physics, Indian Institute of Technology Madras,
Chennai, 600036, India.}

\preprint{IITM/PH/TH/2004/8}

\begin{abstract} The entanglement dynamics of spin chains is investigated
using Heisenberg-XY spin Hamiltonian dynamics.
The various measures of two-qubit entanglement are calculated analytically
in the time-evolved state starting from initial states with no entanglement and 
exactly one pair of maximally-entangled qubits. The localizable entanglement
between a pair of qubits at the end of chain captures the essential features
of entanglement transport across the chain, and it displays the difference
between an initial state with no entanglement and an initial state with
one pair of maximally-entangled qubits.
\vspace{1pc}
\end{abstract}
\maketitle

Quantum entanglement in spin systems is an extensively-studied field in
recent years, in the advent of growing realization that
entanglement can be a resource for quantum information
processing\cite{Nielsen,Peres}. Within this general field,
entanglement of spin-${1\over2}$ degrees of freedom, qubits, has
been in focus for an obvious reason of their paramount importance
for quantum computers, not to mention their well-known
applicability in various condensed-matter systems, optics and
other branches of physics. For a pure state of many qubits,
quantum entanglement is a measure of how a subsystem is correlated
to the rest of the system, which can be quantified by the von
Neumann entropy of the reduced density matrix of the subsystem.

For a system of a large number of qubits, how different pairs of qubits
are entangled in a pure state cannot easily be specified, even
with known diagonal and off-diagonal correlation functions. There are
many pairwise entanglement measures of pure quantum states
of many qubits, as we shall discuss below. Essentially, any two-qubit
entanglement measure depends
on the reduced density matrix, $\rho_{ij}$ of the marked pair of qubits $i$ and
$j$. The reduced density matrix is obtained from the original many-qubit pure
state $|\psi\rangle$ through a partial trace over the rest of the qubits,
$\rho_{ij}= Tr_{[ij]}  {|\psi\rangle}\langle{\psi}|$. Using the two-qubit basis $|{00}
\rangle,
|{01}\rangle,|{10}\rangle,|{11}\rangle$, the reduced density matrix has the form 
\begin{equation}\rho_{ij}=\left(\begin{array}{cccc}
        A_{ij}&E_{ij} &F_{ij}&G_{ij}\\
        E_{ij}^{\star}& B_{ij}&H_{ij}&I_{ij}\\
        F_{ij}^{\star}&H_{ij}^{\star}&C_{ij}&J_{ij}\\
        G_{ij}^{\star}&I_{ij}^{\star}&J_{ij}^{\star}&D_{ij} \end{array} \right).\end{equation}

The two-qubit reduced density matrix above represents a mixed state in general,
though we started with a pure many-qubit state.
We can calculate von Neumann entropy from the eigenvalues of $\rho_{ij}$,
which quantifies how
these two sites are entangled with the rest of the
system. However, a measure of
entanglement between these two sites cannot be straightforwardly
quantified, because of the mixed-state structure.
In the above each of the matrix elements can be expressed as an expectation
value of a two-qubit operator in the initial state $|{\psi}\rangle$. Using the
Pauli operators for a qubit, $\sigma^+|{0}\rangle=|{1}\rangle,\sigma^+|{1}\rangle=0,
\sigma^z|{0}\rangle=-|{0}\rangle,\sigma^z|{1\rangle}=|{1}\rangle.$,
for instance,
\begin{equation}
H_{ij}=\langle{01}|\rho_{ij}|{10}\rangle=\langle{\psi}|\sigma_i^+\sigma_j^-|{\psi}\rangle,
\end{equation}
and similarly the other matrix elements are expectation values of appropriate
operators.
A good measure of impurity of the 
two-qubit state is $M_{ij}= 1 - Tr \rho_{ij}^2$, which is zero if $\rho_{ij}$
represents a pure state of the given pair of qubits.
In general the
reduced density matrix represents a mixed state, giving a nonzero value
for $M_{ij}$.
The von Neumann entropy calculated as
$-Tr \rho_{ij}  \log \rho_{ij}$ quantifies entanglement of this
pair with the rest of the qubits. When $M_{ij}=0$, the reduced density matrix
represents a pure state for the pair of qubits and the von Neumann entropy
would be zero. For $M_{ij}\ne 0$, the von Neumann entropy will be nonzero,
indicating an entanglement of the pair with the rest of the system, and 
possibly a mutual entanglement among the pair of qubits. 
The pairwise concurrence\cite{Wootters} is a
measure that quantifies the mutual entanglement of the marked pair of qubits. 
It carries the information of entanglement sharing between
site $i$ and any other site. If the concurrence between $i$ and
$j$ is one, would imply neither of these two sites is entangled
with the rest of the system, and hence a pure-state assignment can
be done to these sites.
If the concurrence less than one, that
would imply site $i$ is sharing entanglement with site $j$ and the
rest of the system too.
The concurrence measure\cite{Wootters}
is given as
\begin{equation}
{\cal C}_{ij}=~{\rm max}~ (0, \lambda_1^{1/2}-\lambda_2^{1/2}-\lambda_3^{1/2}-
\lambda_4^{1/2}).
\end{equation}
In the above $\lambda_i$ are the eigenvalues in decreasing order of the
matrix $\rho_{ij} \hat \rho_{ij}$, where $\hat \rho_{ij}$ is the time-reversed
matrix,
$\hat \rho_{ij}=\sigma_i^y \otimes \sigma_j^y \rho_{ij}^* \sigma_i^y \otimes
\sigma_j^y$.
The above concurrence measure captures the essential information of
mutual entanglement of the marked pair of qubits. 
However, it is not clear if a given value
of the concurrence for a pair of qubits would imply that one can separate
the pair of qubits in an entangled state, and use the entanglement as a
resource operationally for quantum information processing.

The localizable entanglement (LE) between
two qubits of a many-qubit state is the maximal amount of entanglement 
system that can be concentrated on the given pair of qubits, on an average, by doing a
measurement on the rest of the qubits \cite{Cirac}.

The LE between two qubits can be isolated, by an appropriate measurement
on the rest of the qubits, and thus can be used as resource. Also, this
has a direct bearing on the correlations present between the pair of qubit
in the parent many-qubit state. The calculation of LE could be very tedious,
as one has to search through every possible measurement basis for the rest
of the qubits. However, there are useful upper and lower bounds on LE,
that are calculable from the diagonal correlation functions alone \cite{Cirac}.
Let us denote
\begin{equation}
Q_{ij}^x\equiv \langle \sigma_i^x \sigma_j^x\rangle - 
\langle \sigma_i^x\rangle
\langle \sigma_j^x\rangle,
\end{equation}
and similarly $Q_{ij}^y,Q_{ij}^z$ in terms of $yy$ and $zz$ correlation
functions, and
\begin{equation}
s_{ij}^{\pm}\equiv (1\pm 
\langle \sigma_i^z \sigma_j^z\rangle)^2 - 
(\langle \sigma_i^z\rangle\pm
\langle \sigma_j^z\rangle)^2.
\end{equation}
In terms of the above correlation functions, the bounds on LE are
\begin{equation}
\mbox{max }(|Q_{ij}^x|, |Q_{ij}^y|, |Q_{ij}^z|) \le LE_{ij} \le {
\sqrt{s_{ij}^+}+\sqrt{s_{ij}^-}\over 2}.
\end{equation}
Now, each of the matrix elements of the reduced density matrix
$\rho_{ij}$ involves correlation functions. After some manipulations
using the properties of the Pauli matrices, and their expectation
values in the initial state, we have $Q^x=2 \mbox{Re}(H+G)-4
\mbox{Re}(E+J) \mbox{Re}(I+F), Q^y=2 \mbox{Re}(H-G)-4
\mbox{Im}(E+J)\mbox{Im}(I+F), Q^z=4 (AD-BC)$, and $s_{ij}^+=16 AD,
s_{ij}^-=16 BC$. Here the $ij$ subscripts have been suppressed for
clarity.
 
The entanglement properties depend only
on the structure of entanglement of a many-body state
without reference to a Hamiltonian. However, here we would like to
investigate the entanglement dynamics of a simple interacting spin
systems on a lattice.  Consider a spin-${1\over2}$ spin system, a
nearest-neighbour anisotropic Heisenberg model in one dimension
with a Hamiltonian
\begin{equation}
H=  {K\over2} \sum_i(S_i^+ S_{i+1}^- +{\rm H.C.}) + \Delta S_i^z
S_{i+1}^z - E_0\end{equation} where $S_i^+$ is spin-raising operator at
site $i$, $K$ is the exchange interaction strength, and $\Delta$ a measure of 
anisotropy. A constant, $E_0$ the energy of a state with all down spins or
all up spins, has been added on for convenience. Working in a
diagonal basis of $S_i^z$ for every site $i$, there are two states
per site, $viz$., an up-spin state (denoted by $|1\rangle$ and a down-spin 
state denoted by $|0\rangle$). A
many-spin state can be characterized by the number of down spins,
as the total $S^z$ is a good quantum number. 
We confine ourselves here to spin systems in one dimension, as eigenstates
are known in this case through Bethe-Ansatz, and are well studied.  

The concurrence measure of entanglement in the eigenstates of the
above model have been investigated\cite{Zanardi,Arul,Subrah1}, and the
entanglement dynamics have been
discussed\cite{Bose,Subrah2,Osborne}. The localizable entanglement has
been studied for the ground state\cite{Korepin}. In this study we will
focus on the dynamics of the various entanglement measures for the
above model.  Our strategy is to start from a given initial state
(either with no entangled pairs of qubits or with exactly one pair of
qubits maximally entangled and the rest in an unentangled state), and
study its entanglement properties using the measures discussed above,
through the time evolution using Heisenberg dynamics.  The initial
state $|\psi(0)\rangle$, let us say with exactly one entangled pair of
qubits at one end of the chain, will evolve into a state
$|\psi(t)\rangle$ at a later time $t$, with a complicated distribution
of entanglement, with dynamically changing correlation functions. We
shall investigate how entanglement transport takes place across the
spin chain, by studying the entanglement of a pair of qubits at the
other end of the chain.
\begin{figure}
\label{FIG1}
\includegraphics{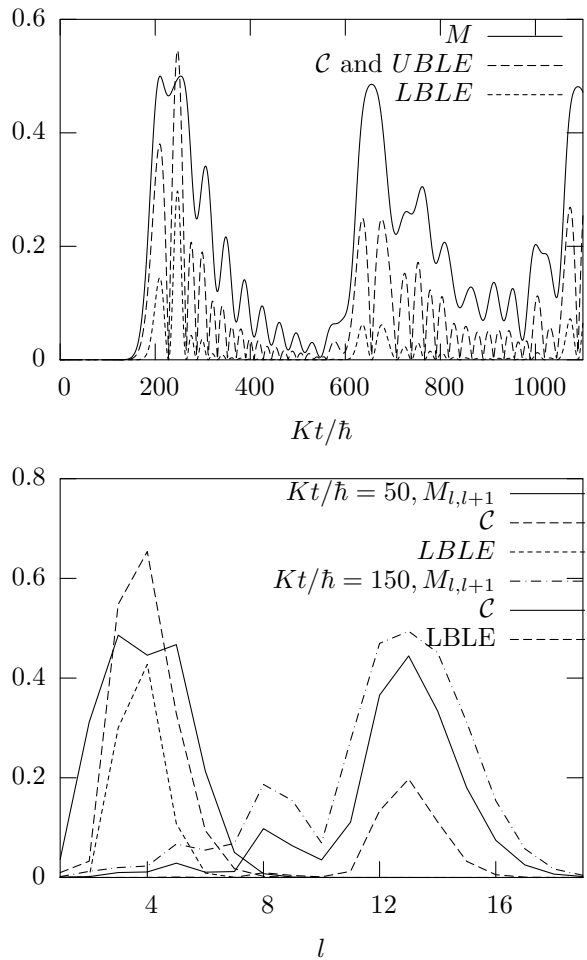}
\caption{The entanglement measures for the last pair of qubits in
an unentangled initial state $|\psi_u\rangle$ are
plotted vs. time, for a finite open chain of $N=20$ qubits. The upper bound
on LE coincides with the concurrence measure in this sector with exactly
on down spin. In the second figure the entanglement for nearest neighbour
qubits $l,l+1$ are plotted vs $l$ for two different times $Kt/\hbar=50,150$.}
\end{figure}

The dynamics of Heisenberg Hamiltonian above conserves the number of down spins
in a given state. This implies that, 
for an initial state with a definite number for down spins, at future
times the reduced density matrix for a pair of qubits can have nonzero diagonal
elements 
and a nonzero off-diagonal element $H_{ij}$ in Eq.~(1), and we have
\begin{equation}
E_{ij}=F_{ij}=G_{ij}=I_{ij}=J_{ij}=0,
\end{equation}
which simplifies the expressions for the entanglement measures substantially.
We have for a many-qubit state with a definite number of down spins (number
of qubits in state $|0\rangle$),
\begin{eqnarray}
M_{ij}=&1-A_{ij}^2 -B_{ij}^2 -C_{ij}^2 -D_{ij}^2-2 |H_{ij}|^2,\\
{\cal C}_{ij}=& 2 \,\mbox{max}\, (0,|H_{ij}|-
\sqrt{A_{ij}D_{ij}}),\\
LE_{ij}\ge& 2\,(\sqrt{A_{ij}D_{ij}}+
 \sqrt{B_{ij}C_{ij}}),\\
LE_{ij} \le&  \mbox{max} \, (4 |A_{ij}D_{ij}-B_{ij}C_{ij}|, 2 \mbox{Re} H_{ij}).
\end{eqnarray}
A further simplification occurs for a state with exactly one down spin. Here,
$A_{ij}$ the first diagonal element of the reduced density matrix vanishes.

Let us first consider an initial state with no entangled qubit pairs, given
as
\begin{equation}
|\psi_u(t=0)\rangle = |01111...\rangle.
\end{equation}
Under the time evolution with the above Hamiltonian, the down spin located
at the first site propagates, through spin-1 magnon excitations. Denoting
a state with the down spin at site $l$ by $|l\rangle$, the one-magnon
eigenstates are given as $|q\rangle=\sum_l \phi_l(q)|l\rangle$. Here the
one-magnon eigenfunction for a chain of $N$ qubits with open boundary
conditions is
\begin{equation}
\phi_l(q)= \sqrt{2\over N+1} \sin(ql), q={\pi n\over N+1}, n=1,..N,
\end{equation}
and the one-magnon eigenvalues are $\epsilon(q)=K\cos{q}$. The interaction
term in the Hamiltonian does not figure in the magnon excitations in this
sector with exactly one down spin. Now the time-evolved
state at a time $t$ can be written as
\begin{equation}
|\psi_u(t)\rangle = \sum_l \gamma_{l,1}(t) |l \rangle
\end{equation}
where the time-dependent wave function is given by
\begin{equation}
\gamma_{l,m}= \sum_q e^{-i\epsilon_q t} \phi_l(q) \phi_m^{\star}(q).
\end{equation}
The matrix elements of the reduced density matrix can be expressed in terms
of the function $\gamma_{l,m}$. Let us consider the pair of qubits located
at sites $i$ and $j$  We have
$B=|\gamma_{i,1}|^2,C=|\gamma_{j,1}|^2,D=1-B-C,H=\gamma_{i,1}\gamma_{j,1}^
{\star}$.

\begin{figure}
\label{FIG2}
\includegraphics{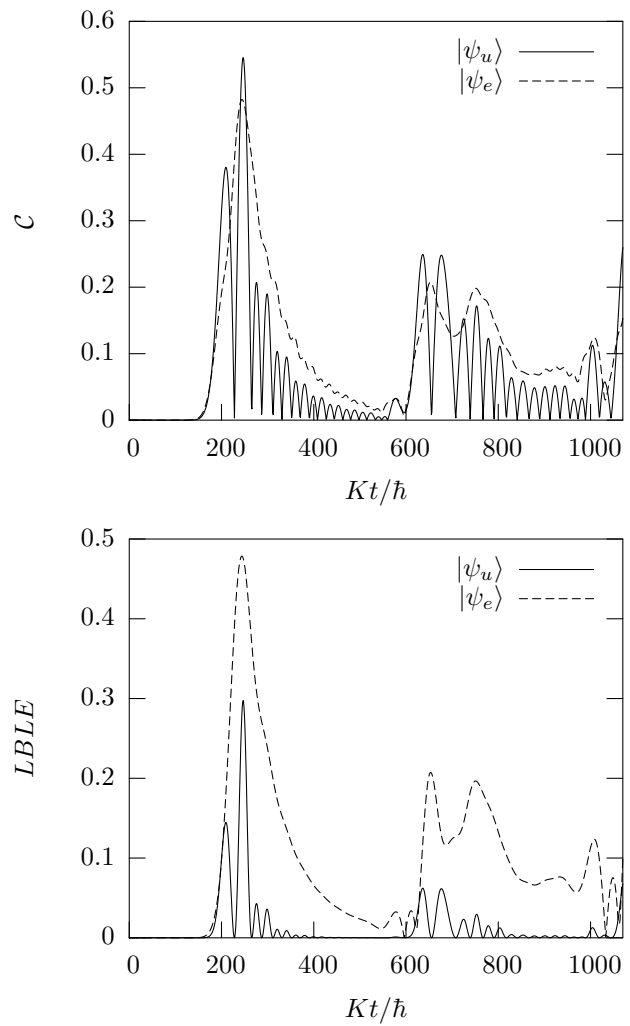}
\caption{The concurrence and LE  measures of entanglement are plotted 
against time for
unentangled and entangled initial states for the last pair of qubits for
an open chain of $N=20$ qubits.}
\end{figure}
Though we started with a state with no entanglement, in time pairs of qubits
get entangled due to propagation of the down spin through magnon excitations.
The different entanglement measures have been plotted in Fig.~1 as functions of 
time for the pair of qubits at the end of the chain (i.e. $i=N-1,j=N$) for
a chain of $N=20$ qubits with open boundary conditions.
The entanglement
measures show time-dependent oscillatory behaviour, owing to the sinusoidal
wave function. One interesting point is that the concurrence equals the upper bound
on LE for one-magnon states. 

The impurity measure, $M_{ij}$ is largest for
almost all time, as it has contributions from two types of entanglements, namely
the entanglement of this pair with the rest of the chain and the mutual
entanglement between the two qubits. The concurrence is for the most part smaller than the
impurity measure, but it is always greater than or equal to the lower bound on
LE. The entanglement measures for various nearest neighbour pairs
of qubits are shown for two different times, $Kt/\hbar=50,150$. All the measures
show peaks for some pair of qubits, the peak shifting with the
time. 

It is important to note that all the measures have similar overall structures,
for instance when the impurity of the pair of qubits is a local maximum, it is
also
the time when the pair is maximally entangled. Thus although purity (or lack of it) is
not an entanglement measure, even in simple models as these there is a strong correlation 
between them. The LE as well as the concurrence share this property and hence we may 
say that usable entanglement is created with time at the end of the chain, even as the
qubits become more mixed.

Let us now turn our attention to an initial state with the first pair of
qubits maximally entangled, and the rest of qubits in state $|1\rangle$, given
as
\begin{equation}
|\psi_e(t=0)\rangle = {1\over \sqrt{2}}|(01+10) 11111 \rangle.
\end{equation}
The above state still has a definite number of down spins. Analogous to the
unentangled state, we can write the state at a later time in this case as
\begin{equation}
|\psi_e(t)\rangle = {1\over \sqrt{2}}\sum_l (\gamma_{l,1}+\gamma_{l,2})|l\rangle.
\end{equation}
Now the nonzero elements of the reduced density matrix are given by $D=1-B-C$,
and
\begin{eqnarray}
B_{ij}=&{1\over 2}|\gamma_{i,1}+\gamma_{i,2}|^2,\\
C_{ij}=&{1\over 2}|\gamma_{j,1}+\gamma_{j,2}|^2,\\
H_{ij}=&{1\over 2}(\gamma_{j,1}^{\star}+\gamma_{j,2}^{\star})(\gamma_{i,1}+
\gamma_{i,2}).
\end{eqnarray}
The envelope of the entanglement measures for the last pair of qubits as functions of
 time show similar behaviour as that of the unentangled state $|\psi_u\rangle$ shown
in Fig.~1. We plot the different measures for the two cases of unentangled and
entangled initial states in Fig.~2. The concurrence of the last two qubits 
for the initial state $|\psi_e\rangle$ is smoother than the case when the initial state is 
is totally unentangled, namely $|\psi_u\rangle$, in which case the entanglement shows rapid
oscillations that repeatedly nearly vanish.
The lower bound on LE, ($LBLE$) shows
up the difference between the two cases, even in the envelope. A part of the entanglement in LE
for the last pair of qubits shown in the figure can be ascribed to the 
entanglement generated through dynamics, as in the case of the unentangled
state, but a substantial part comes from the transport of initial entanglement
of the first pair of qubits, through the Heisenberg dynamics. 

In Fig.~3 we show a density plot of (the lower bound of) the localizable
 entanglement between neighbouring qubits,$LE_{l,l+1}(t)$, as a function of $l$, the location of the pair and
the time, for a chain of $N=20$ qubits for an initial state with the first
pair of qubits in a maximally-entangled state, that is $|\psi_e\rangle$.
 The brighter regions in the plot correspond to larger entanglement
for the given pair at the given time. We can see the front of bright ``entanglement'' patch
moving across the chain and reach the end linearly in time, getting reflected, and again moving
back and forth. After one traversal the bright patch is not so clearly seen, as entanglement begins
to be shared by more than one neighbouring pair of qubits, simultaneously.
For the first passage across the chain, the location of the patch at time $t$
is, $l=vt$, where $v=K$ is the velocity of the front. For a nonzero 
entanglement at a distance $l$, the pair of qubits located at $l,l+1$ have
to first become a part of a mixed state. This happens through the transport of
the down spin located at the first site via one-magnon excitations. The time
scale associated with the magnons is $\tau=1/K$, and $v=1/\tau$.
\begin{figure}
\label{FIG3}
\includegraphics*{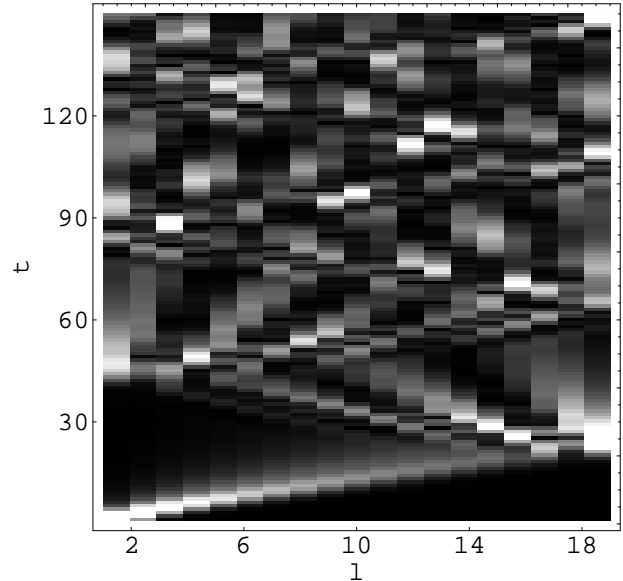}
\caption{$LBLE_{l,l+1}(t)$ is plotted as function of $l$ (along x-axis)
and the time $t$, with brighter regions corresponding to larger entanglement,
for a chain of $N=20$ qubits for the initial state $|\psi_e\rangle$. The
bright patch travels across the chain to reach the end.}
\end{figure}

In conclusion, we
have investigated the dynamics of Heisenberg-XY spin chain using the
various pairwise entanglement measures. Initial states with no entangled
pairs of qubits, and with exactly one pair of entangled qubits are studied
through the time-evolution, and the their dynamics of entanglement is juxtaposed.
We have shown that the localizable entanglement bound captures the essential
features of entanglement transport, and contrasts better the
dynamics of
a state with no initially-entangled pair of qubits from the dynamics of a
state with a pair of qubits maximally entangled. The entanglement transport
occurs through magnon excitations, and the first passage to the end of the
chain of the pair entanglement front occurs with a uniform rate.

\end{document}